\documentclass[aps,prl,twocolumn,superscriptaddress,showpacs]{revtex4-1}

\usepackage{amsmath}
\usepackage{amssymb}
\usepackage{graphicx}
\usepackage{dcolumn}
\usepackage{bm}
\usepackage{hyperref}
\usepackage{epsfig}
\usepackage{rotating}

\begin{document}

\title{Distinct high-$T$ transitions in underdoped Ba$_{1-x}$K$_{x}$Fe$_{2}$As$_{2}$}

\author{R. R. Urbano}
\affiliation{National High Magnetic Field Laboratory, Florida State University, Tallahassee, Florida 32306-4005, USA.}
\author{E. L. Green}
\affiliation{National High Magnetic Field Laboratory, Florida State University, Tallahassee, Florida 32306-4005, USA.}
\author{W. G. Moulton}
\affiliation{National High Magnetic Field Laboratory, Florida State University, Tallahassee, Florida 32306-4005, USA.}
\author{A. P. Reyes}
\affiliation{National High Magnetic Field Laboratory, Florida State University, Tallahassee, Florida 32306-4005, USA.}
\author{P. L. Kuhns}
\affiliation{National High Magnetic Field Laboratory, Florida State University, Tallahassee, Florida 32306-4005, USA.}
\author{E. M. Bittar}
\affiliation{Instituto de F\'{\i}sica "Gleb Wataghin", Unicamp, 13083-970, Campinas-S\~{a}o Paulo, Brazil.}
\author{C. Adriano}
\affiliation{Instituto de F\'{\i}sica "Gleb Wataghin", Unicamp, 13083-970, Campinas-S\~{a}o Paulo, Brazil.}
\author{T. M. Garitezi}
\affiliation{Instituto de F\'{\i}sica "Gleb Wataghin", Unicamp, 13083-970, Campinas-S\~{a}o Paulo, Brazil.}
\author{L. Bufai\c{c}al}
\affiliation{Instituto de F\'{\i}sica "Gleb Wataghin", Unicamp, 13083-970, Campinas-S\~{a}o Paulo, Brazil.}
\author{P. G. Pagliuso}
\affiliation{Instituto de F\'{\i}sica "Gleb Wataghin", Unicamp, 13083-970, Campinas-S\~{a}o Paulo, Brazil.}

\date{\today}

\begin{abstract}

In contrast to the \emph{simultaneous} structural and magnetic
first order phase transition $T_{0}$ previously reported, our
detailed investigation on an underdoped
Ba$_{0.84}$K$_{0.16}$Fe$_{2}$As$_{2}$ single crystal
unambiguously revealed that the transitions are not concomitant. The tetragonal ($\tau$: $\emph{I4/mmm}$) - orthorhombic ($\vartheta$: $\emph{Fmmm}$) structural transition
occurs at $T_{S}\simeq$ 110 K, followed by an adjacent
antiferromagnetic (AFM) transition at $T_{N}\simeq$ 102 K. Hysteresis
and coexistence of the $\tau$ and $\vartheta$ phases over
a finite temperature range observed in our NMR experiments confirm the first order character of
the structural transition and provide evidence that both
$T_{S}$ and $T_{N}$ are strongly correlated. Our data also show
that superconductivity (SC) develops in the $\vartheta$ phase
below $T_{c}$ = 20 K and coexists with long range AFM. This new
observation, $T_{S}\neq T_{N}$, firmly establishes another
similarity between the hole-doped BaFe$_{2}$As$_{2}$ via K
substitution and the electron-doped iron-arsenide superconductors.

\end{abstract}


\maketitle
The novel class of iron-based high-$T_{c}$ superconductors $\emph{Ln}$OM$\emph{Pn}$ ($\emph{Ln}$: Lanthanides; M: Mn, Fe, Co, and Ni; $\emph{Pn}$: P and As) has
been one of the most highly investigated superconducting materials
since their discovery \cite{Kamihara, Chen_Nature}. Their
appearance has revitalized the field of superconductivity (SC),
providing another window for exploring the interplay between
magnetism and unconventional SC. These superconductors are
characterized by AFM correlations and long
range ordering throughout the phase diagram, often coexisting with
SC deep into the superconducting dome
\cite{Inosov_NatPhys2010, Drew_NatMat2009, ParkPRL102,
PrattPRL103, AndyPRL103}. The parent (magnetic) undoped compound
usually becomes superconducting upon either chemical
(electron- or hole-) doping or hydrostatic pressure. As a
consequence, a dramatic suppression of the AFM order is commonly
observed. The K-doped BaFe$_{2}$As$_{2}$ (Ba-122) is the
first reported oxygen-free iron-arsenide superconductor with $T_{c}$
as high as 38 K for the optimal doping $x_{K}\simeq 0.4$
\cite{Rotter_PRL101_2008}. Their ThCr$_{2}$Si$_{2}$-type crystal structure possesses a Fe-As layer similar to that of the $\emph{Ln}$FeAs(O$_{1-x}$F$_{x}$)
\cite{Kamihara} but not suffering from synthesis issues
associated with the control of oxygen or fluorine.

BaFe$_{2}$As$_{2}$ is reported to undergo a
simultaneous structural ($\tau$ - $\vartheta$) and AFM transitions
at $T_{0}\simeq 138$ K \cite{Rotter_PRL101_2008,
Huang_PRL101_2008, ChenEPL85, NiPRB78, Kitagawa_JPSJ77_2008}.
Neutron diffraction measurements in the ordered state revealed an
ordered moment of 0.87$\mu_{B}$ at the Fe site with an ordering
wave-vector $\mathbf{q}$ = (1,0,1) \cite{Huang_PRL101_2008}, later
confirmed to be a commensurate AFM state by zero-field $^{75}$As
NMR experiments \cite{Fukazawa_JPSJ77}. Coexistence between
itinerant AFM and SC has been reported for a wide
range of potassium doping, $0.1\lesssim x_{K} \lesssim 0.3$
\cite{Rotter_AC_2008, ChenEPL85} and is a common feature
in other iron-arsenide compounds  \cite{PrattPRL103, LesterPRB79,
ProzorovPRB80, NiPRB78, Kasahara_cond-mat}. However, until now,
the reports regarding the structural/magnetic transition in the
Ba-122 family show a conflicting behavior: when BaFe$_{2}$As$_{2}$
is doped with electrons via "in plane" Co or Ni substitution of Fe
ions, the temperature of the first order $\tau - \vartheta$ transition and
the AFM ordered phase is monotonically suppressed and the transition separates into two, with the $\tau - \vartheta$ transition occurring at $T_{S} > T_{N}$, similar to what is found for
$\emph{Ln}$OM$\emph{Pn}$ compounds
\cite{McGuirePRB78, PrattPRL103, LesterPRB79, ProzorovPRB80, ChuPRB79}.
However, the separation of the structural and AFM transitions has
not been observed for the hole-doping case via "out of
plane" K substitution of Ba and, to the best of our knowledge, no experimental
evidence of $T_{S} \neq T_{N}$ has been reported, even though they present similar phase diagrams \cite{Rotter_AC_2008, ChenEPL85}. Since the AFM
state is strongly coupled to the lattice distortions generated by
the $\tau - \vartheta$ transition, further
investigation is needed to resolve this issue and unravel
the interplay between lattice and spin degrees of freedom. NMR is a suitable microscopic technique which can shed new light on the transitions, complementing the bulk
measurements. In this letter we address the problem using a single crystal of underdoped Ba$_{0.84}$K$_{0.16}$Fe$_{2}$As$_{2}$. We demonstrate for the first time that $T_{S} \neq T_{N}$ for the hole-doped Ba$_{1-x}$K$_{x}$Fe$_{2}$As$_{2}$ and suggest that the electronic and magnetic properties of this material share remarkable similarities with the electron-doped iron-arsenides such as
Ba(Fe$_{1-x}$\emph{M}$_{x}$)$_{2}$As$_{2}$ (\emph{M} = Co, Ni).

Single crystals of Ba$_{1-x}$K$_{x}$Fe$_{2}$As$_{2}$ were
grown using Sn-flux method. The previously
reported Sn-flux growth method \cite{NiPRB78b} was modified as described in Ref. \onlinecite{Sn-flux_Growth_method}. The data reported here suggest that our growth method leads to higher quality crystals with very small or negligible Sn-contamination than previously reported. Powder X-ray diffraction on the ground crystals shows single phase with no trace of FeAs impurities as
suggested for polycrystalline samples. The dc susceptibility measurements $\chi_{dc}(T, H)$ were
performed in a Quantum Design Physical Property Measurement System (PPMS) Vibrating Sample Magnetometer (VSM) with zero field cooled (ZFC) and field cooled (FC) procedures with $H\bot c$
= 20 Oe. In-plane electrical resistivity $\rho_{el}(T)$ was
measured in a PPMS using the standard four-probe method with $H\bot i_{\textrm{ab}}$. The specific heat $C_{p}(T)$ measurements were performed on a PPMS with the thermometer and addenda properly calibrated in field. The same crystal used in the bulk measurements was mounted on a NMR probe equipped with a goniometer, which allowed a fine alignment of the crystallographic axes with the external field. The field-swept $^{75}$As NMR spectra ($I$ = 3/2; $\gamma/2\pi$ = 7.2919 MHz/T) were obtained by stepwise summing the Fourier transform of the spin-echo signal.
\begin{figure}[!ht]
\vspace{-3mm}
\epsfig{figure=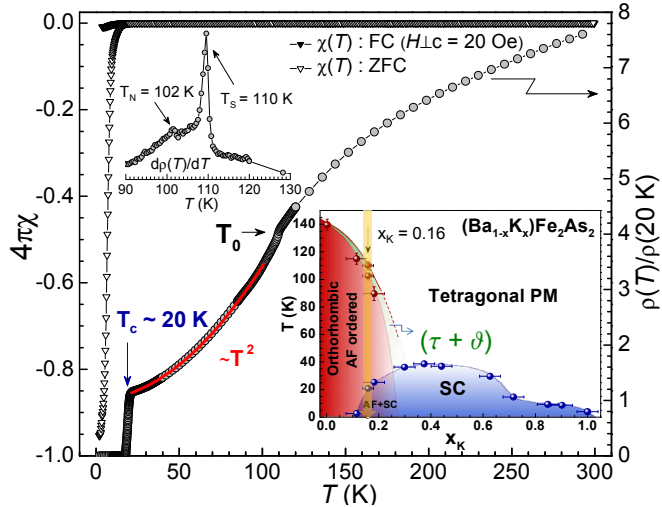, angle=90, scale=0.32}
\vspace{-3mm}
\caption{(color online) $T$-dependence of $\rho_{el}(T)$ for Ba$_{0.84}$K$_{0.16}$Fe$_{2}$As$_{2}$. The solid line is a fit of $\rho(T) = \rho_{0} + A~T^{2}$ with $\rho_{0} = (287.9 \pm 0.3)\mu\Omega$cm and $A = (66.4\pm 0.1)$n$\Omega$cm. Top inset: d$\rho(T)$/d$T$ revealing that $T_{S} \neq T_{N}$. The FC and ZFC measurements of $\chi_{dc}(T)$ with $H\perp c$ = 20 Oe for the same crystal is also shown. Bottom inset: phase diagram of Ba$_{1-x}$K$_{x}$Fe$_{2}$As$_{2}$ adapted from \cite{Rotter_AC_2008}. The novel coexisting orthorhombic PM phase ($\tau + \vartheta$) is highlighted. The arrow indicates our sample in the phase diagram.}
\end{figure}

The temperature dependence of the in-plane $\rho_{el}(T)$ of our Ba$_{0.84}$K$_{0.16}$Fe$_{2}$As$_{2}$
single crystal is shown in Fig. 1. We observe a pronounced
anomaly around $T_{0}\simeq 105$ K where $\rho_{el}(T)$ changes its
\emph{concavity}, not observed as large as that before for the
underdoped material \cite{Rotter_NJP_2009, Rotter_PRL101_2008,
WangPRB78, LuoPRB80}. As shown in the inset of
Fig. 1, d$\rho (T)$/d$T$ demonstrates that the $\tau - \vartheta$ structural phase transition occurs at $T_{S}\simeq$ 110 K, followed by an AFM transition at $T_{N}\simeq$ 102 K.  Our sample exhibits a sharp superconducting transition at $T_{c}\simeq 20$ K with
$\Delta T_{c}\lesssim 1.5$ K. $T_{c}$ is reduced by only 2 K in
fields of $\sim$ 9 T, reflecting the large $H_{c2}$ in these compounds \cite{HQYuan_Nature457}. The residual resistivity ratio RRR $\equiv
\rho$(300 K)/$\rho$(20 K) $\simeq 7.8$ and the extrapolated
residual resistivity $\rho_{0}$ = (287.9 $\pm$ 0.3)$\mu\Omega$.cm
are in good agreement with the self-flux grown
K-doped BaFe$_{2}$As$_{2}$ \cite{Rotter_NJP_2009,
Rotter_PRL101_2008, WangPRB78, LuoPRB80}. These features
indicate the high homogeneity and good quality of our single
crystal and demonstrate that the Sn-flux technique can yield iron-arsenide samples with high quality. Previous reports on Ba-122 grown in Sn-flux have stated that the Sn impurities suppress the AFM ordering
temperature by nearly 40\%, increases resistivity with decreasing
the temperature below $T_{N}$, broadens the
transition and alters the critical dynamics
\cite{SH_PRB78_2008}. On the other hand, the
resistivity of our sample does not
follow this behavior and continuously decreases with
decreasing temperature as shown in Fig. 1. Further, the observation of
$T_{S} \neq T_{N}$ along with the sharp $T_{c}$ is a
strong evidence that our sample shows negligible or no Sn-incorporation.

The magnetic susceptibility measured on the same
single crystal are also presented in Fig. 1. FC ($H\bot c = 20$ Oe) and ZFC cycles are shown as a function of temperature. The ZFC $\chi_{dc}$ data
shows SC onset at $T_{c}\simeq 20$ K and a diamagnetic response corresponding to
nearly 98\% of superconducting volume. The FC curves
reveal a much smaller diamagnetic response commonly observed for the underdoped Ba-122. The \emph{anomalies} in $\rho_{el}(T)$ data around $T_{0}$ are only observed in $\chi_{dc}$ with fields $H\gtrsim 1$ T (not shown), with temperature dependence similar to that observed for BaFe$_{2}$As$_{2}$ \cite{Rotter_PRB78_2008}.
\begin{figure}[!hb]
\vspace{-3mm}
\epsfig{figure=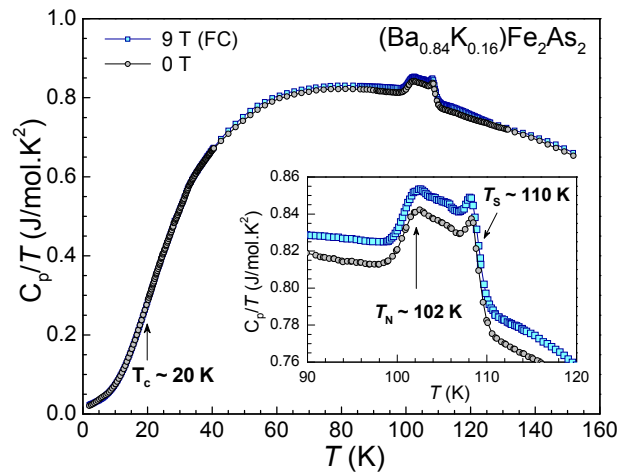, angle=90, scale=0.29}
\vspace{-3mm}
\caption{(color online) $T$-dependence of $C_{p}/T$ for Ba$_{0.84}$K$_{0.16}$Fe$_{2}$As$_{2}$. Inset:
blow up of the $T_{S}$ and $T_{N}$ transitions. Arrows indicate the onset of the $\tau - \vartheta$ transition at $T_{S}\simeq 110$ K and the AFM long range order at $T_{N}\simeq$ 102 K.}
\end{figure}

In Fig. 2 we show the temperature dependence of the specific
heat $C_{p}/T$ plotted as a function of temperature for $H$ = 0 and 9 T for the same single crystal. $C_{p}/T$ is essentially field independent except near $T_{S}$ and $T_{N}$ which indicates that the magnetic field might have some influence over the conservation of the entropies associated with each transition.

In contrast to the single anomaly observed previously in the $C_{p}$ of K-doped Ba-122 samples
\cite{Rotter_NJP_2009}, we observe two distinct peaks occurring at slightly different temperatures consistent with d$\rho (T)$/d$T$. We associate the sharper peak at $T_{S}\simeq$ 110 K with the first order $\tau - \vartheta$ transition
and the broader one at $T_{N}\simeq$ 102 K as the AFM
transition. This new observation for the underdoped
Ba$_{1-x}$K$_{x}$Fe$_{2}$As$_{2}$ is consistent with the
double transitions observed in electron-doped Ba-122, indicating higher homogeneity and crystallographic quality of our sample \cite{ProzorovPRB80, LesterPRB79, ChuPRB79}. It has been suggested that the splitting is due to either, the formation of fluctuating AFM domains below $T_{S}$
which become pinned at lower $T_{N}$ or, an Ising transition at $T_{S}$ followed by AFM order at $T_{N}$
\cite{Mazin_Nat.Phys5, Qi_PRB80_2009}. Both scenarios involve
coupling between lattice and magnetic fluctuations. $C_{p}/T$ shows no visible
anomaly at $T_{c}\simeq 20$ K where bulk SC is
evident and inferred from Fig. 1. This behavior agrees with
a previous report on the underdoped
Ba$_{1-x}$K$_{x}$Fe$_{2}$As$_{2}$ with $x$ = 0.1 and 0.2
\cite{Rotter_NJP_2009}. However, we were able to identify the
maximum of a broad bump around 20 K from the derivative d$C_{p}(T)$/d$T$ (not
shown). The peak defining $T_{c}$ in the $C_{p}$ data is only evident for optimally doped samples ($x_{K}\simeq$ 0.4) when long range AFM order is fully suppressed \cite{Rotter_NJP_2009}. We also point out that the transition temperatures observed by $\chi_{dc}$, $\rho_{el}(T)$ and $C_{p}/T$ data are
consistent with the phase diagram with x$_{K}$ = 0.16 (see Fig. 1).
\begin{figure}[!h]
\vspace{-3mm}
\epsfig{figure=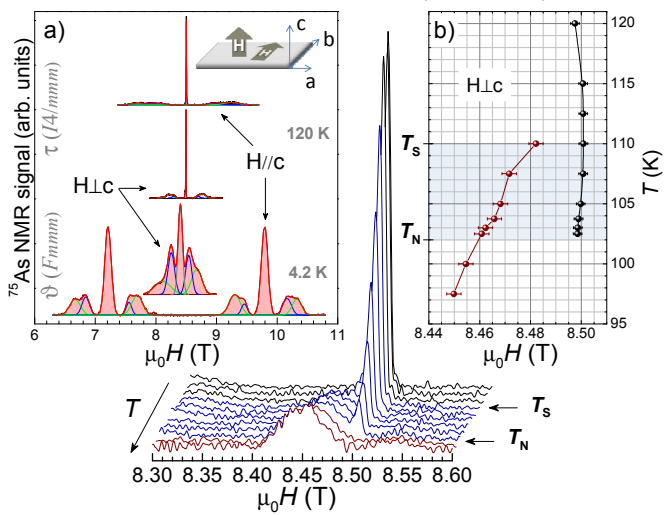, angle=90, scale=0.32}
\vspace{-3mm}
\caption{(color online) $^{75}$As NMR spectra of Ba$_{0.84}$K$_{0.16}$Fe$_{2}$As$_{2}$ at 120 K (PM state) and 4.2 K (AFM+SC coexisting state) for $H\| c$ and $H\bot c$. In the PM state, the spectra shows a sharp central transition $(+1/2 \leftrightarrow -1/2)$ at 8.5 T with broad satellite peaks from the $(\pm 1/2 \leftrightarrow \pm 3/2)$ transitions. Below $T_{N}$, the $H_{\textrm{int}}\|c$ symmetrically splits the NMR spectra for $H\|c$. For $H\bot c$, a shift is observed \cite{Kitagawa_JPSJ77_2008}. The main panel presents the central transition with $H\bot c$ where $\tau - \vartheta$ phase coexistence is evident; b) Peak position of the lines obtained from gaussian fits.}
\end{figure}

Motivated by the possible coupling between the AF order and the
structural distortion we have also performed a microscopic
investigation by NMR experiments. Fig. 3(a) presents the
$^{75}$As-NMR spectra obtained by sweeping up $H\bot c$ at constant frequency $\nu$ = 62.238
MHz. The spectra present the typical features of a
nuclear spin $I = 3/2$ with Zeeman and quadrupolar couplings. The
spectra at 120 K represents the normal paramagnetic (PM) state
while the broad spectra at 4.2 K belong to the coexisting
AFM+SC states.

In the PM state, the spectra show a sharp central line with a full width at half maximum (FWHM) of
36 kHz ($\sim$ 50 Oe) and two sets of broader satellite lines split by the quadrupolar interaction of the As nuclei with the local electric field gradient (EFG). The single and sharp central transition is a microscopic evidence for a high quality single crystal. The satellites are split due to the strain introduced by doping, the different number of Ba neighbors to As sites, and the broadening is due to a distribution of both EFG and hyperfine fields. The quadrupole splitting is proportional to the EFG at the As, $V_{zz}$, through the relation: $\nu_{Q} = eQV_{zz}/2\sqrt{1+\eta^{2}/3}$, where Q is the nuclear quadrupole moment of $^{75}$As and $\eta$ is the asymmetry parameter of the EFG ($\eta$ = 0 for the $\tau$ phase). $V_{zz}$ arises from the hybridization between the As-4$p$ and the Fe-3$d$ orbitals with an added contribution from any non-cubic distribution of surrounding ions, resulting in two distinct sets of quadrupole satellite lines through doping with different quadrupole frequencies, $\nu_{Q}$ = 4.07(1) and 2.94(1) MHz. These are larger than $\nu_{Q}$ = 2.21 MHz observed in the undoped system \cite{Kitagawa_JPSJ77_2008} and agree with previous observations that $\nu_{Q}$ increases with increased doping \cite{Julien_EPL,Mukuda_PhysC469}. It also confirms that this sample has electronic but not chemically (or macroscopically) segregated inhomogeneities \cite{ParkPRL102}. The twinned structural domains are only observed in the orthorhombic ($\vartheta$) ordered phase \cite{Kitagawa_JPSJ77_2008} and may be responsible for the line broadening below $T_{N}$.

Below the AFM ordering temperature, internal hyperfine fields $H_{\textrm{int}}$ develop along the $c$-axis generated by the commensurate AF ordered Fe moments at the As. These $H_{\textrm{int}}$ split the $^{75}$As-NMR spectra doubling the number of resonance lines when
$H\|c$ (Fig. 3(a)). The lines broaden
due to a distribution of EFG and static hyperfine fields produced by the
lattice distortions at the $\tau - \vartheta$ transition. The internal
field $H_{\textrm{int}}^{\| c}$(4.2 K) $\simeq$ 1.29(1) T is obtained from the splitting of the central lines (after taking the second order quadrupolar shift $\nu_{Q}^{2}/\nu_{0}$ and the demagnetization contribution into account), since the
resonance field is given by $H_{\textrm{eff}} = H_{0} \pm
H_{\textrm{int}}$. This agrees with the only NMR report so far in the underdoped regime, $H_{\textrm{int}}$ = 1.23(5)T on a $x$ = 0.3 aligned powder sample taken at zero field \cite{Fukazawa_JPSJ78}. However, these values are smaller than $H_{\textrm{int}}\simeq 1.48(2)$ T for BaFe$_{2}$As$_{2}$ \cite{Kitagawa_JPSJ77_2008, Fukazawa_JPSJ78, Mukuda_PhysC469}. It has been suggested that the $H_{\textrm{int}}$ decreases with increasing K concentration \cite{Fukazawa_JPSJ78}. Since
$H_{\textrm{int}}$ is a direct measure of the magnetic order parameter, this result is consistent with the suppression of $T_{N}$ and the emergence of SC upon increasing K-doping as shown by the phase diagram in Fig. 1.

Since $H_{\textrm{int}}\|c$, when $H\bot c$, the resonance field is given by the magnitude of the vector sum of the mutually orthogonal external and internal fields, $H_{\textrm{eff}} = \sqrt{H^{2} + H_{\textrm{int}}^{2}}$, causing the shift of the unsplit $^{75}$As-NMR spectra towards lower fields as displayed in Fig. 3(a) \cite{Kitagawa_JPSJ77_2008}. The main panel of Fig. 3 shows the central transition of the $^{75}$As-NMR spectra with $H\bot c$ at several temperatures above and below $T_{S} \simeq$ 110 K and $T_{N} \simeq$ 102 K. Fig. 3(b) presents the peak position of the lines obtained from gaussian fits. The single narrow central transition for temperatures $T\geq T_{S}$ represents the $^{75}$As-NMR spectra in the $\tau$ phase in the PM state. The spectra for $T \lesssim T_{S}$ reveals a sudden appearance of a broad line at slightly lower fields coexisting with the narrow line for a finite $T$-range. The broad peak grows rapidly, while the sharp peak is suppressed with decreasing temperature. Also the peak intensity of the spectrum revealed hysteresis around $T_{S}$. We define this temperature as the onset of the first order structural transition $\tau - \vartheta$ coexisting with the emergence of the broad central transition of the orthorhombic PM phase. The sharp line completely disappears at $T_{N}$ and only the broad line of the AF ordered phase remains below this temperature. The entire spectrum broadens and is shifted towards lower fields as a consequence of the pronounced change of the symmetry of the EFG and the ordered Fe moments generating the $H_{\textrm{int}}^{\|c}$ at the $^{75}$As nuclei. The spectrum with $H\| c$ taken at 4.2 K supports the absence of the $\tau$ phase below $T_{N}$ (see Fig. 3(a)). This observation is a microscopic evidence that the transitions are strongly coupled. It could be a consequence of the stress and/or pressure at the boundaries of the twinned domains due to the orthorhombic distortion and thus, lowering $T_{N}$. Such effect could lead to a non-monotonic behavior of the order parameter but with a monotonic variation in temperature as suggested by a recent neutron scattering study on the related SrFe$_{2}$As$_{2}$ compound \cite{Li_cond-mat}. H. Li and coworkers claim that the structural transition is probably an order-disorder one, based on the fact that they observed some residual $\tau$ phase below the transition $T_{S}$ and some $\vartheta$ above it. Similar behavior is also found in Ba(Fe$_{0.953}$Co$_{0.047}$)$_{2}$As$_{2}$ \cite{PrattPRL103}. In contrast to this observation, the NMR data in Fig. 3 reveal a complete disappearance of the sharp peak associated with the $\tau$ phase below $T_{N}\simeq$ 102 K and show no traces of the broad peak associated with the twinned $\vartheta$ ordered phase above $T_{S}\simeq$ 110 K, attesting that the remnant minor structural phases are absent or negligible in our sample. This result also suggests that either, sample dependence (quality) is probably an issue in iron-arsenide compounds or, the order-disorder scenario is a particular feature of the SrFe$_{2}$As$_{2}$ compound. A persistent secondary NMR line was observed in the $^{75}$As NMR spectra of BaFe$_{2}$As$_{2}$ within the tetragonal PM phase and was attributed to an impurity phase containing As \cite{SH_PRB78_2008}.

Regarding the coexistence of the orthorhombic AFM phase with SC suggested by the bulk measurements, we conclude from our NMR investigation that they coexist microscopically in underdoped compositions, at least on a lattice parameter length scale as inferred from the spin-spin and spin-lattice relaxation rates, $1/T_{2}$ and $1/T_{1}$, respectively (not shown). In addition, we notice little or no change in the $^{75}As$ NMR spectra while crossing $T_{c}$ and we observe the broad spectra associated with the $\vartheta$ structure within the SC phase, with no traces of the $\tau$ phase, confirming the crystallographic homogeneity of the sample. This clearly indicates that the $^{75}$As NMR signal is predominantly from the AF regions even below the SC transition. Since the $H_{int}$ is a direct measure of the magnetic order parameter, its suppression by roughly 15\% suggests that the ordered Fe moment must be decreasing with addition of potassium and provides evidence that the AFM order might be related with the SC. The results reported here demonstrate that coupling to the lattice plays an important role in the AFM transition and may be a remarkable signature within the underdoped region of the phase diagram of the iron-arsenide superconductors.

We thank L. Balicas and N. J. Curro for relevant discussions. Work at NHMFL was performed under the auspices of the NSF through the Cooperative Agreement No. DMR-0654118 and the State of Florida. Work at Unicamp was supported by Fapesp, CNPq and Capes.


\begin{thebibliography}{100}
\bibitem{Sn-flux_Growth_method} Elemental Ba, K, Fe, and As were added to Sn in the
ratio of [(Ba/K)Fe$_{2}$As$_{2}$: Sn = 1:20 and placed in a 5
ml Al$_{2}$O$_{3}$ crucible which was covered with quartz wool
and sealed in an evacuated quartz ampoule. The ampoule
was heated to 850$^{\circ}$C for 0.5h and cooled at 5$^{\circ}$C/h down
to 500$^{\circ}$C. After 48h of in-fux annealing at 500$^{\circ}$C, the
crystals were separated from the Sn-fux in a centrifuge.
\end{thebibliography}
\end{document}